\documentclass[letterpaper, 11pt]{article}
\title
{Disproof of the Neighborhood Conjecture with Implications to SAT}
\author{Heidi Gebauer
\thanks{Institute of
Theoretical Computer Science, ETH Zurich, CH-8092 Switzerland. Email:
gebauerh@inf.ethz.ch. Research is supported by the SNF Grant 200021-118001/1.} 
}

\date{}
\usepackage{amsmath, amsfonts}
\usepackage{amsthm}
\usepackage{subfigure}
\usepackage[dvips]{graphicx} 
\usepackage{fullpage} 

\newcommand{\fref}[1]{Figure~\ref{#1}}

\begin{document}
\bibliographystyle{plain}
\maketitle
\newtheorem{theo}{Theorem} [section]
\newtheorem{defi}[theo]{Definition}
\newtheorem{lemm}[theo]{Lemma}                                                                                                                                                                                                                                                                                                                                                                                                                                                                                                                   
\newtheorem{obse}[theo]{Observation}
\newtheorem{prop}[theo]{Proposition}
\newtheorem{coro}[theo]{Corollary}
\newtheorem{rem}[theo]{Remark}
\newtheorem{opprob}{Open Problem}

\newcommand{\whp}{{\bf whp}}
\newcommand{\prob}{probability}
\newcommand{\rn}{random}
\newcommand{\rv}{random variable}
\newcommand{\hpg}{hypergraph}
\newcommand{\hpgs}{hypergraphs}
\newcommand{\subhpg}{subhypergraph}
\newcommand{\subhpgs}{subhypergraphs}
\newcommand{\bH}{{\bf H}}
\newcommand{\cH}{{\cal H}}
\newcommand{\cT}{{\cal T}}
\newcommand{\cF}{{\cal F}}
\newcommand{\cG}{{\cal G}}
\newcommand{\cD}{{\cal D}}
\newcommand{\cC}{{\cal C}}

\newcommand{\ideg}{\mathsf {ideg}}
\newcommand{\lv}{\mathsf {lv}}
\newcommand{\nga}{n_{\text{game}}}
\newcommand{\avdaneg}{\overline{\deg}}
\newcommand{\ed}{e_{\text{double}}}

\newcommand{\danger}{\mathsf {dang}}
\newcommand{\avdanan}{\overline{\danger}}

\newcommand{\degb}{\deg_{B}}
\newcommand{\degm}{\deg_{M}}

\newcommand{\avd}{\overline{D}}

\newcommand{\vbl}{\mathsf {vbl}}

\newcommand{\depth}{\mathsf {depth}}

\newcommand{\de}{\overline{\mathsf {deg}}}

\newcommand{\Tres}{T_{\mathsf {res}}}

\newcommand{\dl}{d_{\mathsf {leaf}}}


\begin{abstract}
We study a Maker/Breaker game described by Beck. As a result we disprove a conjecture of Beck on positional games, establish a connection between this game and SAT and construct an unsatisfiable $k$-CNF formula with few occurrences per variable, thereby improving a previous result by Hoory and Szeider and showing that the bound obtained from the Lov\'{a}sz Local Lemma is tight up to a constant factor. 

The Maker/Breaker game we study is as follows.
Maker and Breaker take turns in choosing vertices from a given $n$-uniform hypergraph $\cal{F}$, with Maker going first.
Maker's goal is to completely occupy a hyperedge and Breaker tries to avoid this. 
Beck conjectures that if the maximum neighborhood size of $\cal{F}$ is at most $2^{n - 1}$ then Breaker has a winning strategy. We disprove this conjecture by establishing an $n$-uniform hypergraph with maximum neighborhood size $3 \cdot 2 ^{n - 3}$ where Maker has a winning strategy. Moreover, we show how to construct an $n$-uniform hypergraph  with maximum degree $\frac{2^{n - 1}}{n}$ where Maker has a winning strategy.

In addition we show that each $n$-uniform hypergraph with maximum degree at most $\frac{2^{n - 2}}{en}$ has a proper halving 2-coloring, which solves 
another open problem posed by Beck related to the Neighborhood Conjecture. 

Finally, we establish a connection between SAT and the Maker/Breaker game we study. We can use this connection to derive new results in SAT. A \emph{$(k,s)$-CNF formula} is a boolean formula in conjunctive normal form where every clause contains exactly $k$ literals and every variable occurs in at most $s$ clauses. The $(k,s)$-SAT problem is the satisfiability problem restricted to $(k,s)$-CNF formulas. Kratochv\'{i}l, Savick\'y and Tuza showed that for every $k \geq 3$ there is an integer $f(k)$ such that every $(k, f(k))$-formula is satisfiable, but $(k, f(k) + 1)$-SAT is already NP-complete (it is not known whether $f(k)$ is computable). 
Kratochv\'{i}l, Savick\'y and Tuza also gave the best known lower bound $f(k) = \Omega\left(\frac{2^{k}}{k}\right)$, which is a consequence of the Lov\'{a}sz Local Lemma.
We prove that, in fact, $f(k) = \Theta\left(\frac{2^{k}}{k}\right)$, improving upon the best known upper bound $O\left((\log k) \cdot \frac{2^{k}}{k}\right)$ by Hoory and Szeider. 
\end{abstract}


\section{Introduction}

A \emph{hypergraph} is a pair $(V,E)$, where $V$ is a finite set whose elements are called \emph{vertices} and $E$ is a family of subsets of $V$, called \emph{hyperedges}. 
We study the following Maker/Breaker game.
Maker and Breaker take turns in claiming 
one previously unclaimed vertex of a given $n$-uniform hypergraph $\cal{F}$, with Maker going first.
Maker wins if he claims all vertices of some hyperedge of $\cal{F}$, otherwise Breaker wins. We say that Maker uses a \emph{pairing strategy} if after claiming his first vertex he divides all but at most one of the remaining vertices of $\cal{F}$ into pairs 
and whenever Breaker claims one vertex of a pair he takes the other one. 

Let $\cal{F}$ be an $n$-uniform hypergraph. The \emph{degree} $d(v)$ of a vertex $v$ is the number of hyperedges containing $v$ and the \emph{maximum degree} $\Delta(\cal{F})$ of a hypergraph $\cal{F}$ is the maximum degree of its vertices. The \emph{neighborhood} $N(e)$ of a hyperedge $e$ is the set of hyperedges of $\cal{F}$ which intersect $e$ and the \emph{maximum neighborhood size} of $\cal{F}$ is the maximum of $|N(e)|$ where $e$ runs over all hyperedges of $\cal{F}$.

The famous Erd\H{o}s-Selfridge Theorem \cite{ESe} states that for each $n$-uniform hypergraph $\cal{F}$ with less than $2 ^{n - 1}$ hyperedges Breaker has a winning strategy. This upper bound on the number of hyperedges is best possible as the following example shows. 
Let $T$ be a rooted binary tree with $n$ levels and let $\cal{G}$ be the hypergraph whose hyperedges are exactly the sets $\{v_{0}, \ldots v_{n - 1}\}$ such that $v_{0}, v_{1}, \ldots, v_{n - 1}$ is a path from the root to a leaf.
Note that the number of hyperedges of $\cal{G}$ is $2^{n - 1}$. 
To win the game on $\cal{G}$ Maker can use the following strategy. In his first move he claims the root $m_{1}$ of $T$. Let $b_{1}$ denote the vertex occupied by Breaker in his subsequent move. 
In his second move Maker claims the child $m_{2}$ of $m_{1}$ such that $m_{2}$ lies in the subtree of $m_{1}$ not containing $b_{1}$. 
More generally, in his $i$th move Maker selects the child $m_{i}$ of his previously occupied node $m_{i - 1}$ such that the subtree rooted at $m_{i}$ contains no Breaker's node.
Note that such a child $m_{i}$ always exists since the vertex previously claimed by Breaker is either in the left or in the right subtree of $m_{i-1}$ (but not in both!). Using this strategy Maker can achieve to own some set $\{v_{0}, \ldots, v_{n - 1}\}$ of vertices such that $v_{0}, v_{1}, \ldots, v_{n - 1}$ is a path from the root to a leaf, which corresponds to some hyperedge of $\cal{G}$. Hence Maker has a winning strategy on $\cal{G}$.

Note that both the maximum neighborhood size and the maximum degree of $\cal{G}$ are $2^{n - 1}$, thus equally large as the number of hyperedges of $\cal{G}$. 
This provides some evidence that in order to be a Maker's win a hypergraph must have largely overlapping hyperedges. Moreover, Beck \cite{B} conjectured that the main criterion for whether a hypergraph is a Breaker's win is not the cardinality of the hyperedge set but rather the maximum neighborhood size, i.e. the actual reason why each hypergraph $\cal{H}$ with less than $2^{n-1}$ edges is a Breaker's win is that the maximum neighborhood of $\cal{H}$ is smaller than $2^{n-1}$.
\newline
\newline
\textbf{Neighborhood Conjecture} (Open Problem 9.1(a), \cite{B}) Assume that $\cal{F}$ is an $n$-uniform \hpg, and its maximum neighborhood size is smaller than $2^{n - 1}$. Is it true that by playing on $\cal{F}$ Breaker has a winning strategy?
\newline
\newline
Further motivation for the Neighborhood Conjecture is the well-known Erd\H{o}s-Lov\'{a}sz 2-coloring Theorem -- a direct consequence of the famous Lov\'{a}sz Local Lemma -- which states that every $n$-uniform hypergraph with maximum neighborhood size at most $2^{n-3}$ has a proper 2-coloring. 
An interesting feature of this theorem is that the board size does not matter.
In this paper we prove by applying again the Lov\'{a}sz Local Lemma that in addition every  $n$-uniform hypergraph with maximum neighborhood size at most $\frac{2^{n-4}}{n}$ has a so called \emph{proper halving} 2-coloring, i.e., a proper 2-coloring in which the number of red vertices and the number of blue vertices differ by at most 1 (see Theorem \ref{theo:NeighbourhoodPartProperHalvingIsOk} for details). This guarantees the existence of a course of the game at whose end Breaker owns at least one vertex of each hyperedge and thus is the winner. 
Hence it is a priori not completely impossible that Breaker has a winning strategy.

In our first theorem we prove that the Neighborhood Conjecture, in this strongest of its forms, is not true, even if we require Maker to use a pairing strategy.

\begin{theo} \label{theo:NeighbourhoodPartanotwork}
There is an $n$-uniform hypergraph $\cal{H}$ with maximum neighborhood size $2^{n-2} + 2^{n - 3}$ where Maker has a winning pairing strategy.
\end{theo}
In his book \cite{B} Beck also poses the following weakening of the Neighborhood Conjecture.
\begin{opprob}  (Open Problem 9.1(b), \cite{B}) \label{opprob:OpenProbSec}
 If the Neighborhood Conjecture is too difficult (or false) then how about if the upper bound on the maximum neighborhood size is replaced by an upper bound $\frac{2^{n - c}}{n}$ on the maximum degree where $c$ is a sufficiently large constant?
\end{opprob}
In the hypergraph $\cal{H}$ we will construct to prove Theorem \ref{theo:NeighbourhoodPartanotwork}
one vertex has degree $2^{n-2}$, which is still high. However, the existence of vertices with high degree is not crucial. 
We also establish a hypergraph with maximum degree $\frac{2^{n - 1}}{n}$ on which Maker has a winning strategy. 
In this case the maximum neighborhood size is at most $2^{n - 1} - n$, which is weaker than Theorem \ref{theo:NeighbourhoodPartanotwork} but also disproving the Neighborhood Conjecture.

\begin{theo} \label{theo:NeighbourhoodPartbnotworkc1}
If $n$ is a sufficiently large power of 2 there is an $n$-uniform hypergraph with maximum degree $\frac{2^{n-1}}{n}$ where Maker has a winning pairing strategy.
\end{theo}
The bound in Theorem \ref{theo:NeighbourhoodPartbnotworkc1} is not tight. Indeed, we can prove the following
\begin{theo} \label{theo:NeighbourhoodPartbnotworkforsomeotherc}
Let $c = \frac{64}{63}$. For every sufficiently large $n$ with $cn$ being a power of 2 there is an $n$-uniform hypergraph with maximum degree $\frac{2^{n-1}}{cn}$ where Maker has a winning pairing strategy.
\end{theo}

Note that by Theorem \ref{theo:NeighbourhoodPartbnotworkc1} the answer to Open Problem \ref{opprob:OpenProbSec}  for $c = 1$ is no. Since the proof of Theorem \ref{theo:NeighbourhoodPartbnotworkc1} 
contains several technical lemmas and long calculations we first establish a slightly weaker construction revealing one of the main ideas of the proof.
\begin{theo} \label{theo:NeighbourhoodPartbweakerclaim}
For every $n \geq 4$ there is an $n$-uniform hypergraph with maximum degree $\frac{2^{n + 2}}{n}$ where Maker has a winning pairing strategy.
\end{theo}

In his book \cite{B} Beck also poses several further weakenings of the Neighborhood Conjecture. The last one is as follows.
\begin{opprob} (Open Problem 9.1(f), \cite{B}) \label{opprb:BeckLatest}
 How about if we just want a proper halving 2-coloring?
\end{opprob}
 It is already known that the answer to Open Problem \ref{opprb:BeckLatest} is positive if the maximum degree is at most $\left(\frac{3}{2} - o(1)\right)^{n}$. According to Beck \cite{B} the real question is whether or not $\frac{3}{2}$ can be replaced by 2. We prove that the answer is yes.

\begin{theo} \label{theo:NeighbourhoodPartProperHalvingIsOk}
For every $n$-uniform hypergraph $\cal{F}$ with maximum degree at most $\frac{2^{n - 2}}{en}$ there is a proper halving 2-coloring.
\end{theo}

\paragraph{Connection to SAT}

Our results also have implications to SAT. Following the standard notation we denote by  \emph{$(k,s)$-CNF} the set of boolean formulas $\cal{F}$ in conjunctive normal form where every clause of $\cal{F}$ has exactly $k$ distinct literals and each variable occurs in at most $s$ clauses of $\cal{F}$. Moreover, we denote by $(k,s)$-SAT the satisfiability problem restricted to formulas in $(k,s)$-CNF.
Tovey \cite{T} proved that every (3,3)-CNF formula is satisfiable but (3,4)-SAT is NP-complete. Hence  $(3,s)$-SAT is trivial for $s \leq 3$, and NP-complete for $s \geq 4$.  
Kratochv\'{i}l, Savick\'y and Tuza \cite{KST} generalized this result by showing that for every $k \geq 3$ there is some integer $s = f(k)$ such that
\begin{itemize}
\item[(i)]{every $(k,s)$-CNF formula with $s \leq f(k)$ is satisfiable, and}
\item[(ii)]{$(k,s + 1)$-SAT is already NP-complete.}
\end{itemize}
For positive integers $k$ the function $f$ can be defined by the equation
\begin{displaymath}
f(k) := \max\{s: \text{every $(k,s)$-CNF formula is satisfiable}\}
\end{displaymath}
The best known lower bound for $f(k)$, a consequence of Lov\'{a}sz Local Lemma, is due to Kratochv\'{i}l, Savick\'y and Tuza \cite{KST}.
\begin{theo} \label{theo:KSTlowerbound} {\bf (Kratochv\'{i}l, Savick\'y and Tuza \cite{KST})}
$f(k) \geq \lfloor \frac{2^{k}}{ek} \rfloor$
\end{theo}
From the other side Savick\'y and Sgall \cite{SS} showed that
$f(k) = O(k^{(1 - \alpha)} \cdot \frac{2^{k}}{k})$ where $\alpha = \log_{3} 4 - 1 \approx 0.26$. This was improved by Hoory and Szeider \cite{HSfa} who proved that $f(k) = O((\log k) \cdot \frac{2^{k}}{k})$, which is the best known upper bound. We close the gap between upper and lower bound by showing that $f(k) = \Theta(\frac{2^{k}}{k})$, implying that the lower bound  in Theorem \ref{theo:KSTlowerbound} is asymptotically tight.
To this end we introduce a new function $f_{\mathsf{bal}}$ which bounds $f$ from above. Then we establish an upper bound for $f_{\mathsf{bal}}(k)$, which also serves as an upper bound for $f(k)$.

A $(k,s)$-CNF formula is called \emph{balanced} if every literal occurs in at most $\frac{s}{2}$ clauses. 
Similarly to $f$ we define the function $f_{\mathsf{bal}}$ by the equation
\begin{displaymath}
f_{\mathsf{bal}}(k) := \max\{s: \text{every balanced $(k,s)$-CNF formula is satisfiable}\}
\end{displaymath}
Clearly, $f(k) \leq f_{\mathsf{bal}}(k)$. 
We can show that the lower bound of Theorem  \ref{theo:KSTlowerbound} is best possible up to a factor of $e$.

\begin{theo} \label{theo:UnsatCNFOnlyPowers}
If $k$ is a sufficiently large power of 2 then $f_{\mathsf{bal}}(k) \leq \frac{2^{k}}{k} - 1$. For every sufficiently large $k$ (not necessarily a power of 2) we have $f_{\mathsf{bal}}(k) \leq 2 \cdot \frac{2^{k}}{k} - 1$.
\end{theo}
\noindent
The first part of Theorem \ref{theo:UnsatCNFOnlyPowers} will be deduced from Theorem \ref{theo:NeighbourhoodPartbnotworkc1}. 
%
It is relatively easy to conclude from this proof that for large enough $k$ we have $f_{\mathsf{bal}}(k) \leq r - 1$ for every $r \geq \frac{2^{k}}{k}$ which is a power of 2, implying the second part.
%

By a standard application of the Lopsided Lov\'{a}sz Local Lemma \cite{ESp} Theorem \ref{theo:KSTlowerbound} can be modified as follows.
\begin{theo} \label{theo:KSTbalancedlowerbound}
$f_{\mathsf{bal}}(k) \geq \lfloor \frac{2^{k + 1}}{ek} \rfloor$
\end{theo}
\noindent
This shows that our upper bound in Theorem \ref{theo:UnsatCNFOnlyPowers} is best possible within a factor of $\frac{e}{2}$.

Recently Moser \cite{M} showed that for $s \leq \frac{2^{k - 6}}{k}$ not only every $(k, s)$-CNF has a satisfying assignment but there is also an algorithm computing such an assignment efficiently.
Theorem \ref{theo:UnsatCNFOnlyPowers} proves that this bound is asymptotically tight. Indeed, for some $(k,\frac{2^{k}}{k})$-CNF formulas we can not find a satisfying assignment efficiently, simply because there is none. 

The formula we will construct to prove Theorem \ref{theo:UnsatCNFOnlyPowers} belongs to the class MU(1) of minimal unsatisfiable CNF-formulas $\cal{F}$ where $m({\cal{F}}) - n({\cal{F}}) = 1$ with $m(\cal{F})$ denoting the number of clauses of $\cal{F}$ and $n(\cal{F})$ denoting the number of variables of $\cal{F}$. This is in contrast to the approach of Hoory and Szeider, whose derivation of the previously best known upper bound of $f(k) = O((\log k) \cdot \frac{2^{k}}{k})$ did not go via an MU(1) formula. 
Formulas in MU(1) have been widely studied (see, e.g., \cite{AL}, \cite{DDK}, \cite{KZ}, \cite{K}, \cite{S}). 
It is an open question whether the unsatisfiable CNF-formulas with the smallest possible number of occurrences per variable (i.e. the unsatisfiable $(k, f(k) + 1)$-CNF formulas) are members of MU(1).
Scheder \cite{Sched} showed that for almost disjoint $k$-CNF formulas (i.e. CNF-formulas where any two clauses have at most one variable in common) this is not true, i.e., no almost disjoint unsatisfiable $(k, \tilde{f}(k) + 1)$-CNF formula is in MU(1), with $\tilde{f}(k)$ denoting the maximum $s$ such that every almost disjoint $(k,s)$-CNF formula is satisfiable.

Hoory and Szeider \cite{HSco} considered the function
\newline
$f_{1}(k) := \max\{s: \text{every $(k,s)$-CNF formula in MU(1) is satisfiable}\}$. 
Clearly, $f_{1}(k) \geq f(k)$. 
They investigated further on $f_{1}(k)$, showed that it is computable and determined the exact values of $f_{1}(k)$ up to $k = 9$. However, it is not clear how close $f(k)$ and $f_{1}(k)$ are.
The construction we establish to prove Theorem \ref{theo:UnsatCNFOnlyPowers} implies at least the asymptotic equality of $f(k)$ and $f_{1}(k)$.

\begin{coro} \label{coro:UnsatCNFimpliesf1}
For large enough $k$ we have $f_{1}(k) \leq 2 \cdot \frac{2^{k}}{k}$, implying that $f(k), f_{1}(k) = \Theta(\frac{2^{k}}{k})$. 
Moreover, for infinitely many $k$ we have $f_{1}(k) \leq \frac{2^{k}}{k}$.
\end{coro}
\noindent
Theorem \ref{theo:UnsatCNFOnlyPowers} and Corollary  \ref{coro:UnsatCNFimpliesf1} are a consequence of the following theorem, which establishes a connection between the game we study and SAT.
We denote by a $(k,s)$-hypergraph a $k$-uniform hypergraph with maximum degree at most $s$ where Maker has a winning pairing strategy.
\begin{theo} \label{theo:equivalence}
We have
\begin{itemize}
\item[(i)] if there is a $(k,s)$-hypergraph then there is an unsatisfiable balanced $(k,2s)$-CNF formula, and
\item[(ii)] if there is an unsatisfiable $(k,s)$-CNF formula then there is a $(k,s)$-hypergraph.
\end{itemize}
\end{theo}
\noindent
Note that Theorem \ref{theo:UnsatCNFOnlyPowers} follows directly from Theorem \ref{theo:NeighbourhoodPartbnotworkc1} and Theorem \ref{theo:equivalence}.

Instead of the maximum degree we could also consider the maximum neighborhood of a formula. To this end we regard a corresponding analogon of $f(k)$: Let $l(k)$ denote the largest integer such that every $k$-CNF formula with maximum neighborhood size at most $l(k)$ is satisfiable. Recall that the Local Lemma gives that $l(k) \geq \lfloor \frac{2^{k}}{e} \rfloor - 1$. From the other side the ``complete formula'' (i.e. the $k$-CNF formula containing all $2^{k}$ clauses over $V =\{x_{1}, \ldots, x_{k}\}$) shows that $l(k) \leq 2^{k} - 2$. The constructions we establish to prove Theorem \ref{theo:NeighbourhoodPartanotwork} and Theorem \ref{theo:UnsatCNFOnlyPowers} lower this upper bound by a factor of 2 (resp. $\frac{3}{2}$).

\begin{theo} \label{theo:neighborhoodbounds}
 We have
\begin{itemize}
 \item[(i)] $l(k) \leq 2^{k-1} - 1$ for $k$ being a sufficiently large power of 2, and
 \item[(ii)] $l(k) \leq 2^{k-1} + 2^{k - 2}$ for $k \geq 3$
\end{itemize}
\end{theo}

Actually we can slightly improve our upper bounds on $f(k)$ and $l(k)$.

\begin{theo} \label{theo:improvedbounds}
Let $c = \frac{64}{63}$. For every sufficiently large $k$ with $ck$ being a power of 2 we have
\begin{itemize}
\item[(i)]  $f(k) \leq \frac{2^{k - 1}}{ck} - 1$ and
\item[(ii)] $l(k) \leq \frac{2^{k-1}}{c} - 1$ 
\end{itemize}
\end{theo}

\paragraph{Notation} Ceiling and floor signs are routinely omitted whenever they are not crucial for clarity. Throughout this paper $\log$ stands for the binary logarithm.
A \emph{binary tree} is an ordered tree where every node has either two or no children.
Let $T$ be a rooted binary tree. A \emph{path} of $T$ is a sequence of vertices $v_{1}, v_{2}, \ldots, v_{j}$ of $T$ where $v_{k}$ is a child of $v_{k - 1}$ for every $k = 2, \ldots, j$. A \emph{branch} of $T$ is a path starting at the root of $T$ and a \emph{full branch} of $T$ is a path from the root to a leaf. 

We define ${\cH_{T}} = {\cH_{T}}(n)$ as the $n$-uniform hypergraph whose hyperedges are the paths of length $n - 1$ in $T$ ending at a leaf. Let ${\cC}_{n}$ be the set of hypergraphs ${\cH_{T}}$ where every leaf of $T$ has depth at least $n - 1$. 
The hypergraphs we will construct to prove Theorem \ref{theo:NeighbourhoodPartanotwork}, Theorem \ref{theo:NeighbourhoodPartbweakerclaim} and Theorem \ref{theo:NeighbourhoodPartbnotworkc1} all belong to 
${\cC}_{n}$.
Depending on the context we consider a hyperedge $e$ of a hypergraph ${\cH}_{T}$ either as a set or as a path in $T$. So we will sometimes speak of the start or end node of a hyperedge. 

\paragraph{Organization of this paper} In Section \ref{sec:countexneicon} we give a counterexample to the Neighborhood Conjecture in the strongest of its forms by proving Theorem \ref{theo:NeighbourhoodPartanotwork}. In Section \ref{sec:degreghypsmallmaxdeg} we establish more regular counterexample hypergraphs and prove Theorem \ref{theo:NeighbourhoodPartbweakerclaim}, Theorem \ref{theo:NeighbourhoodPartbnotworkc1} and Theorem \ref{theo:NeighbourhoodPartbnotworkforsomeotherc}.
In Section \ref{sec:construunsatformusmanei} we establish a strong connection between the game we study and SAT and prove Theorem \ref{theo:equivalence}, Corollary \ref{coro:UnsatCNFimpliesf1}, Theorem \ref{theo:neighborhoodbounds} and Theorem \ref{theo:improvedbounds}
The proof of Theorem \ref{theo:NeighbourhoodPartProperHalvingIsOk} is relegated to the appendix.

\section{Counterexample to the Neighborhood Conjecture} \label{sec:countexneicon}

The next observation will play a crucial role in this paper.
\begin{obse} \label{obse:NeighborhoodGeneralObservationBranch}
Let $T$ be a binary tree such that every leaf has depth at least $n - 1$. Then 
 Maker has a winning pairing strategy on ${\cH_{T}}$.
\end{obse}
\noindent
This can be seen as follows.
Since by assumption every leaf has distance at most $n- 1$ from the root every full branch of $T$ contains a hyperedge.
The two children of a vertex are called \emph{siblings}. 
The set of non-root nodes of $T$ can be divided into pairs of siblings. By first claiming the root of $T$ and then pairing every node with its sibling Maker can finally achieve some full branch of $T$, which by assumption contains a hyperedge. 

\begin{figure} [!htb]
\centering
\includegraphics[width=0.7\textwidth]{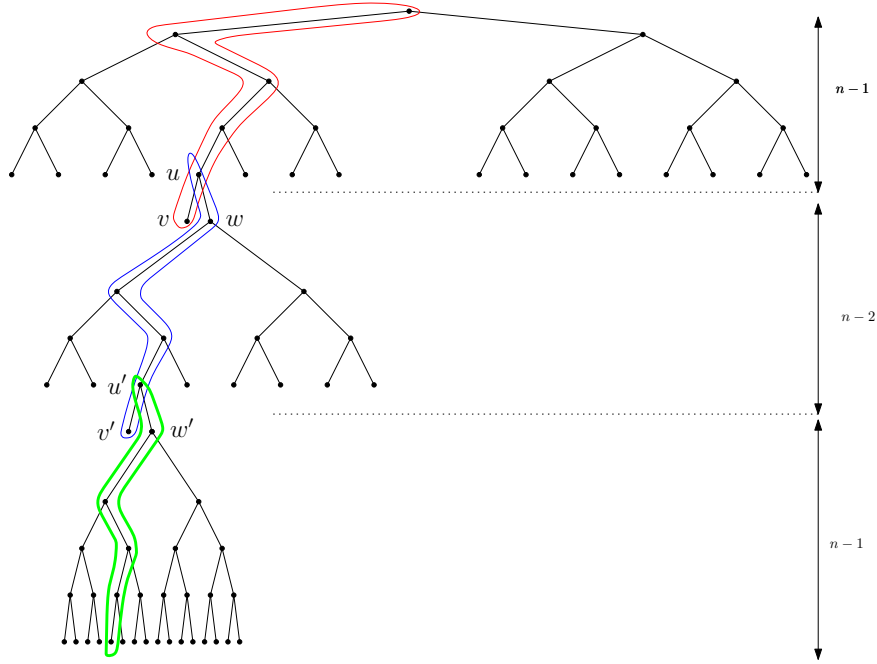}
\caption{An illustration of ${\cal{H}}_{T}$. The marked paths represent exemplary hyperedges.} \label{fig:DisprovingNC}
\end{figure}

\emph{Proof of Theorem \ref{theo:NeighbourhoodPartanotwork}:}  
Due to Observation \ref{obse:NeighborhoodGeneralObservationBranch} it suffices to show the following.
\begin{lemm} \label{lemm:NeighbourhoodPartaAuxilLemm}
There is a binary tree $T$ where every leaf has depth at least $n - 1$ such that ${\cH}_{T}$ has maximum neighborhood size $2^{n - 2} + 2^{n - 3}$.
 \end{lemm}
\noindent 
\emph{Proof:}
Let $T'$ be a full binary tree with $n - 1$ levels. For each leaf $u$ of $T'$ we proceed as follows: We add two children $v$, $w$ to $u$ and let $v$ be a leaf. 
Then we attach a full binary tree $S$ with $n-2$ levels to $w$ (such that $w$ is the root of $S$). 
For each leaf $u'$ of $S$ we  
add two children $v'$, $w'$ to $u'$ and let $v'$ be a leaf. Note that the hyperedge ending at $v'$ starts at $u$. Finally, we attach a full binary $S'$ with $n - 1$ levels to $w'$ (such that $w'$ is the root of $S'$), see \fref{fig:DisprovingNC}. Let $T$ denote the resulting tree.
Clearly, every leaf of $T$ has depth at least $n - 1$.
It remains to show that the maximum neighborhood of ${\cH}_{T}$ is at most $2^{n - 2} + 2^{n - 3}$. \\
\noindent
{\bf Claim:} Every hyperedge $e$ of ${\cH}_{T}$ intersects at most $2^{n - 2} + 2^{n - 3}$ other hyperedges. \\
\noindent
In order to prove this claim, we fix six vertices $u, u', v, v', w, w'$ according to the above description, i.e., $u$ is a node on level $n -2$ whose children are $v$ and $w$, $u'$ is a descendant of $w$ on level $2n - 4$ whose children are $v'$ and $w'$.
Let $e$ be a hyperedge of ${\cH}_{T}$. Note that the start node of $e$ is either the root $r$ of $T$, a node on the same level as $u$ or a node on the same level as $u'$. We now distinguish these cases.
\newline
\newline
{\bf (a)} The start node of $e$ is $r$.
By symmetry we assume that $e$ ends at $v$. According to the construction of $T$ the hyperedge $e$ intersects the $2^{n - 2} - 1$ other hyperedges starting at $r$ and the $2^{n - 3}$ hyperedges starting at $u$. So altogether $e$ intersects  $2^{n - 2} + 2^{n - 3} - 1$ hyperedges, as claimed.
\newline
\newline
{\bf (b)} The start node of $e$ is on the same level as $u$. By symmetry we suppose that $e$ starts at $u$ and ends at $v'$. 
The hyperedges intersecting $e$ can be divided into the following three categories.
\begin{itemize}
\item The hyperedge starting at $r$ and ending at $v$,
\item the $2^{n - 3} - 1$  hyperedges different from $e$ starting at $u$, \enspace and
\item the $2^{n - 2}$ hyperedges starting at $u'$,
\end{itemize}
implying that $e$ intersects at most $2^{n - 2} + 2^{n - 3}$ hyperedges in total.
\newline
\newline
{\bf (c)} The start node of $e$ is on the same level as $u'$. By symmetry we assume that $e$ starts at $u'$. Then $e$ intersects the $2^{n - 2} - 1$ other hyperedges starting at $u'$ and the hyperedge starting at $u$ and ending at $v'$, thus $2^{n - 2}$ hyperedges altogether. \\
\phantom{asfdasfsdfasdfasdfsadfasdfasdf asdfasdfsadfsadf asfasdf asdf asdfasdfsadfasdfadsf asdfasdfasdf as} $\qed$

\section{A Degree-Regular hypergraph with small maximum degree which is a Maker's win.}  \label{sec:degreghypsmallmaxdeg}

Let $T$ be a binary tree where every leaf has depth at least $n - 1$ and let $v$ be a vertex of $T$. Note that the degree of $v$ in ${\cH}_{T}$ equals the number of leaf descendants of $v$ which have distance at most $n - 1$ from $v$.

\subsection{Proof of Theorem \ref{theo:NeighbourhoodPartbweakerclaim}:}
Let $s = \frac{2^{n + 1}}{2^{\lfloor \log n \rfloor}}$ and note that $s \leq \frac{2^{n + 2}}{n}$.
Observation \ref{obse:NeighborhoodGeneralObservationBranch} guarantees that it suffices  to construct a binary tree $T$ where every leaf has depth at least $n - 1$ such that the degree of every vertex in  ${\cH}_{T}$ is at most $s$.
Let $T'$ be a full binary tree of height $n-1$. We subdivide its leaves into intervals of length $\frac{2^{\lfloor \log n \rfloor}}{2}$. Let $\{v_{0}, \ldots, v_{\frac{2^{\lfloor \log n \rfloor}}{2} - 1}\}$ be such an interval. Then we attach a full binary subtree of height $i$ to $v_{i}$. Let $T$ denote the resulting tree. 
It suffices to prove the following.
\begin{prop}
Let $v$ be a vertex of $T$. Then $d(v) \leq s$ in ${\cH}_{T}$.
\end{prop}
\noindent
\emph{Proof:} 
We apply induction on the depth $i$ of $v$.
For $i = 0$ the claim is clearly true. Indeed, the degree of the root is
$\frac{2^{n - 1}}{\frac{2^{\lfloor \log n \rfloor}}{2}} = \frac{2^{n}}{2^{\lfloor \log n \rfloor}} = \frac{s}{2}$.
Now suppose that $v$ has depth $i \in \{1, \ldots, \frac{2^{\lfloor \log n \rfloor}}{2} - 1 \}$.
Note that the set of descendants of $v$ on level $n - 1$ can be subdivided into 
$\frac{2^{n - 1 - i}}{\frac{2^{\lfloor \log n \rfloor}}{2}} \geq 1$ 
intervals. Let $v'$ denote the parent of $v$.
By construction the number of leaf descendants which have distance at most $n - 2$ from $v$ equals 
$\frac{d(v')}{2}$. 
Moreover, every interval $\{v_{0}, \ldots, v_{\frac{2^{\lfloor \log n \rfloor}}{2} - 1}\}$ 
gives raise to $2^{i}$ leaves on level $n - 1 + i$, implying that the number of leaf descendants of $v$ which have distance exactly $n -1$ from $v$ equals 
$\frac{2^{n - 1 - i}}{\frac{2^{\lfloor \log n \rfloor}}{2}} \cdot 2^{i} = \frac{2^{n}}{2^{\lfloor \log n \rfloor}} = \frac{s}{2}$.
So altogether $d(v) \leq \frac{d(v')}{2} + \frac{s}{2} \leq s$. 
It remains to consider the case where $v$ has depth at least $\frac{2^{\lfloor \log n \rfloor}}{2}$. By construction no leaf of $T$ has depth larger than 
$\frac{2^{\lfloor \log n \rfloor}}{2} + n - 2$, implying that the degree of $v$ is at most the degree of its parent. \hfill $\qed$

\subsection{Proof of Theorem \ref{theo:NeighbourhoodPartbnotworkc1}:}  
Let $s = \frac{2^{n - 1}}{n}$. 
Due to Observation \ref{obse:NeighborhoodGeneralObservationBranch} it suffices to prove the following.
\begin{lemm}  \label{lemm:constructioninvolvedtreereduction}
There is a nonempty binary tree $T$ where 
\newline
(i) every leaf has depth at least $n - 1$ and
\newline
(ii) for every vertex $v$ of $T$ the number of leaf descendants which have distance at most $n - 1$ from $v$ is bounded by $s$.
\end{lemm}
\noindent
\emph{Proof:} We need some notation first. 
Let $T$ be a binary tree and let $v$ be a vertex of $T$.
In the following we slightly abuse notation and denote by the \emph{degree} $d(v)$ of $v$ the number of leaf descendants which have distance at most $n - 1$ from $v$. (Note that if some leaves of $T$ have depth smaller than $n - 1$, $d(v)$ might differ from the degree of $v$ in ${\cH}_{T}$.)
Moreover, to every node $w$ of $T$ we assign a \emph{distance-sequence} $D_{w} = (x_{0}, x_{1}, \ldots, x_{n - 1})$ where $x_{i} \cdot \frac{s}{2^{i + 1}}$ is the number of leaf descendants of $w$ which have distance $n - 1 - i$ from $w$.  This notation encodes the degree of $w$ in a weighted fashion, which allows us to describe our most frequent operations in a more compact way. Note that $d(w) = \sum_{i = 0}^{n - 1} x_{i} \cdot \frac{s}{2^{i + 1}}$.
%
\begin{obse} \label{obse:mainsteps}
We have
\begin{list}{\labelitemi}{\leftmargin=0.4em}
\item[(i)] Let $T, T'$ be binary trees whose roots have distance sequence $(x_{0}, \ldots, x_{n - 1})$ and $(x'_{0}, \ldots, x'_{n - 1})$, respectively. Let $v$ be a vertex with left subtree $T$ and right subtree $T'$. Then 
\newline
$D_{v} = (\frac{x_{1} + x'_{1}}{2}, \ldots, \frac{x_{n - 1} + x'_{n - 1}}{2}, 0)$. 
\item[(ii)] Let $T'$ be a binary tree whose root has distance sequence $(x_{0}, \ldots, x_{n - 1})$ and let $T$ be a full binary tree of height $h \leq n - 1$. By attaching a copy of $T'$ to every leaf $l$ of $T$ (such that $l$ is the root of $T'$) we obtain $D_{v} = (x_{h}, \ldots, x_{n - 1}, 0, \ldots, 0)$ for the root $v$ of $T$.
\end{list}
\end{obse}
\noindent
We need some more notation. 
Let $x_{0}, \ldots, x_{n - 1} \in \mathbb{N}$.
A \emph{$(x_{0}, x_{1}, \ldots, x_{n- 1})$-tree} is a nonempty binary tree where every node has degree at most $s$ and $D_{r} = (x_{0}, x_{1}, \ldots, x_{n - 1})$ for the root $r$. A sequence $(x_{0}, \ldots, x_{n - 1})$ is \emph{plausible} if 
$x_{i} \cdot \frac{s}{2^{i + 1}} \in \mathbb{N}$ for every $i$, $i = 0, \ldots, n - 1$. (Clearly, every  sequence $(x_{0}, \ldots, x_{\log s - 1}, 0, \ldots, 0)$ with $x_{0}, \ldots, x_{\log s - 1} \in \mathbb{N}$ is plausible.) 
Note that $D_{v}$ is plausible for every node $v$ of a binary tree. 
To prove Lemma \ref{lemm:constructioninvolvedtreereduction} it suffices to show the following.
\begin{lemm} \label{lemm:NeighborConjeShorterProofforCompResu}
There is an $(x_{0}, 0, 0, \ldots, 0)$-tree for some $x_{0} \geq 0$.
\end{lemm}
\noindent 
Lemma \ref{lemm:NeighborConjeShorterProofforCompResu} guarantees that there is a nonempty binary tree where every vertex has degree at most $s$ and every leaf has depth at least $n - 1$, which implies 
Lemma \ref{lemm:constructioninvolvedtreereduction}. \hfill $\qed$
%
%
%
%

\emph{Proof of Lemma \ref{lemm:NeighborConjeShorterProofforCompResu}:} We divide the proof of Lemma \ref{lemm:NeighborConjeShorterProofforCompResu} into three propositions. Let 
$r = \lfloor \frac{\log s}{2} \rfloor - 1$.

\begin{prop} \label{prop:mainprop1}
There is a 
$(0, \underbrace{2, \ldots, 2}_{\lceil \frac{r}{2}  \rceil}, 0, \underbrace{4, \ldots, 4}_{\lfloor \frac{r}{2}  \rfloor}, 0, \ldots 0)$-tree.
\end{prop}
\begin{prop} \label{prop:mainprop2}
Let $j \leq \lfloor \frac{r}{2} \rfloor - 1$. 
\newline
If there is a 
$(0, \underbrace{2, \ldots, 2}_{r - j - 1}, 0, \underbrace{4, \ldots, 4}_{j + 1}, 0, \ldots 0)$-tree then there is a $(0, \underbrace{2, \ldots, 2}_{r- j}, 0, \underbrace{4, \ldots, 4}_{j}, 0, \ldots 0)$-tree.
\end{prop}
\begin{prop} \label{prop:mainprop3}
Let $i \leq r - 1$. 
\newline
If there is a $(0, \underbrace{2, 2, \ldots, 2}_{i + 1}, 0, \ldots, 0)$-tree then 
there is a 
$(0, \underbrace{2, \ldots, 2}_{i}, 0, \ldots, 0)$-tree.
\end{prop}
\noindent
Note that Proposition \ref{prop:mainprop1} - \ref{prop:mainprop3} together imply Lemma \ref{lemm:NeighborConjeShorterProofforCompResu} (with $x_{0} = 0$). 
Before proving Proposition \ref{prop:mainprop1} - \ref{prop:mainprop3} we first state some general propositions. 
For every distance sequence $(x_{0}, \ldots, x_{n - 1})$ we let $\de(x_{0}, \ldots, x_{n - 1})$ denote the degree of a vertex $v$ with $D_{v} = (x_{0}, \ldots, x_{n - 1})$ divided by $s$, i.e., $\de(x_{0}, \ldots, x_{n - 1}) = \sum_{i = 0}^{n - 1} \frac{x_{i}}{2^{i + 1}}$.
\begin{prop} \label{prop:NeighborConjeHelpProp2}
Let $r \leq \log s$ and let $y_{r}, y_{r + 1}, \ldots, y_{\log s - 1}$ be integers such that
\newline
$\de(y_{r}, \ldots, y_{\log s - 1}, 0, \ldots, 0) \leq 1$. Then (i) $\de(\underbrace{1, \ldots, 1}_{r}, y_{r}, \ldots, y_{\log s - 1}, 0, \ldots, 0) \leq 1$ and 
\newline
(ii) if there is a $(\underbrace{1, \ldots, 1}_{r}, y_{r}, \ldots, y_{\log s - 1}, 0, \ldots, 0)$-tree then there is a $(y_{r}, \ldots, y_{\log s - 1}, 0, \ldots, 0)$-tree.
\end{prop}
\noindent 
\emph{Proof:} 
We first show (ii). 
Let $T'$ be a  $(\underbrace{1, \ldots, 1}_{r}, y_{r}, \ldots, y_{\log s - 1}, 0, \ldots, 0)$-tree and let $T$ be a full binary tree of height $r$. To each leaf $l$ of $T$ we attach a copy of $T'$ (such that $l$ is the root of $T'$). According to Observation \ref{obse:mainsteps} (for $h = r$) we have $D_{v} =  (y_{r}, \ldots, y_{\log s - 1}, 0, \ldots, 0)$ for the root $v$ of $T$. It remains to show that every vertex has degree at most $s$. Let $v_{i}$ be a node on level $i$ of $T$. Note that due to Observation \ref{obse:mainsteps} (for $h = i$) we obtain $D_{v_{i}} = (\underbrace{1, \ldots, 1}_{i}, y_{r}, \ldots, y_{\log s - 1}, 0, \ldots, 0)$. 
\newline
We get
$\de(D_{v_{i}}) = 
\frac{1}{2} + \frac{1}{4} + \ldots + \frac{1}{2^{i}} + 
\sum_{j = 0}^{\log s - 1 -r} \frac{y_{r + j}}{2^{i + 1 + j}} = 
1 - \frac{1}{2^{i}} + 
\frac{1}{2^{i}} \cdot \de(y_{r}, \ldots, y_{\log s - 1}, 0, \ldots, 0) 
\leq 1$.
\newline
The last inequality follows directly from our assumption that $\de(y_{r}, \ldots, y_{\log s - 1}, 0, \ldots, 0) \leq 1$. Hence every vertex of $T$ has degree at most $s$, which concludes the proof of (ii). By inserting $i = r$ in the above proof we immediately obtain (i). \hfill $\qed$

\begin{prop} \label{prop:NeighborConjeHelpProp3}
Let $x_{1}, \ldots, x_{n - 1}, x_{1}', \ldots, x_{n - 1}'$ be integers such that 
 $(0, x_{1}, \ldots, x_{n - 1})$, 
\newline
$(0, x_{1}', \ldots, x_{n - 1}')$, $(\frac{x_{1} + x_{1}'}{2}, \ldots, \frac{x_{n - 1} + x_{n - 1}'}{2}, 0)$ are plausible and $\de(\frac{x_{1} + x_{1}'}{2}, \ldots, \frac{x_{n - 1} + x_{n - 1}'}{2}, 0) \leq 1$. Then
\begin{list}{\labelitemi}{\leftmargin=0.4em}
\item[(i)] $\de(0, x_{1}, \ldots, x_{n - 1})$, $\de(0, x_{1}', \ldots, x_{n - 1}') \leq 1$ \enspace and
\item[(ii)] If there is a $(0, x_{1}, \ldots, x_{n - 1})$-tree and a $(0, x_{1}', \ldots, x_{n - 1}')$-tree
\newline
then there is a $(\frac{x_{1} + x_{1}'}{2}, \ldots, \frac{x_{n - 1} + x_{n - 1}'}{2}, 0)$-tree.
\end{list}
\end{prop}
\noindent
\emph{Proof:} 
(i) follows directly from the fact that $\de(0, x_{1}, \ldots, x_{n - 1}) = \sum_{i = 1}^{n - 1} \frac{x_{i}}{2^{i + 1}} = \sum_{i = 0}^{n - 2} \frac{x_{i + 1}}{2 \cdot 2^{i + 1}}
\leq \sum_{i = 0}^{n - 2} \frac{x_{i + 1} + x_{i + 1}'}{2 \cdot 2^{i + 1}} = \de(\frac{x_{1} + x_{1}'}{2}, \ldots, \frac{x_{n - 1} + x_{n - 1}'}{2}, 0) \leq 1$ (and similarly for $\de(0, x'_{1}, \ldots, x'_{n - 1})$). So it remains to show (ii). Let $T_{1}$ be a $(0, x_{1}, \ldots, x_{n - 1})$-tree and let $T_{2}$ be a $(0, x_{1}', \ldots, x_{n - 1}')$-tree. We take a new node $w$ and attach $T_{1}$ and $T_{2}$ as left and right subtree, respectively and let $T$ denote the resulting tree.
By Observation \ref{obse:mainsteps} $D_{w} = (\frac{x_{1} + x_{1}'}{2}, \ldots, \frac{x_{n - 1} + x_{n - 1}'}{2}, 0)$. Together with the fact that $\de(\frac{x_{1} + x_{1}'}{2}, \ldots, \frac{x_{n - 1} + x_{n - 1}'}{2}, 0) \leq 1$ this implies (ii). \hfill $\qed$
\begin{prop} \label{prop:NeighborConjeHelpProp4}
Let $y_{0}, y_{1}, \ldots, y_{\log s - 1}$ be integers with $\sum_{i = 0}^{\log s - 1} y_{i} \geq 2^{n - \log s}$ such that
\newline
$\de(y_{0}, y_{1}, \ldots, y_{\log s - 1}, 0, \ldots, 0) \leq 1$. Then there is a $(y_{0}, y_{1}, \ldots, y_{\log s - 1}, 0, \ldots, 0)$-tree.
\end{prop}
\noindent
\emph{Proof:} 
Note that a tree consisting of a single node is a $(0, 0, \ldots, 0, 2^{n - \log s})$-tree. By repeatedly applying Proposition \ref{prop:NeighborConjeHelpProp3} we get that there is a 
$(\underbrace{0, \ldots, 0}_{i + n - \log s}, 2^{n - \log s}, \underbrace{0, \ldots, 0}_{\log s - i - 1})$-tree for every $i \leq \log s - 1$. Suppose that there are
$y_{i}$ 
\newline $(\underbrace{0, \ldots, 0}_{i + n - \log s}, 2^{n - \log s}, \underbrace{0, \ldots, 0}_{\log s - i - 1})$-trees for every $i \in \{0, \ldots, \log s - 1\}$. So all in all there are $\sum_{i = 0}^{\log s - 1} y_{i} \geq 2^{n - \log s}$ trees. Applying Proposition \ref{prop:NeighborConjeHelpProp3} $n - \log s$ times shows that there is a $(y_{0}, \ldots, y_{\log s - 1}, 0, \ldots, 0)$-tree. \hfill $\qed$

It remains to show Proposition \ref{prop:mainprop1} - \ref{prop:mainprop3}.

\emph{Proof of Proposition \ref{prop:mainprop1}:} By Proposition \ref{prop:NeighborConjeHelpProp2} it suffices to show that there is a 
\newline
$(\underbrace{1, \ldots, 1}_{\log s - r - 4}, 0, \underbrace{2, \ldots, 2}_{\lceil \frac{r}{2} \rceil}, 0, \underbrace{4, \ldots, 4}_{\lfloor \frac{r}{2} \rfloor}, \underbrace{0, \ldots, 0}_{n - \log s + 2})$-tree. According to Proposition \ref{prop:NeighborConjeHelpProp3} it suffices to show that there exists both a 
\newline
$(\underbrace{0, \ldots, 0}_{\log s - r - 3}, 0, \underbrace{0, \ldots, 0}_{\lceil \frac{r}{2} \rceil + 2 \lfloor \frac{r}{2} \rfloor - \frac{n}{2}}, \underbrace{4, \ldots, 4}_{\frac{n}{2} - 2 \lfloor \frac{r}{2} \rfloor} , 0, \underbrace{8, \ldots, 8}_{\lfloor \frac{r}{2} \rfloor}, \underbrace{0, \ldots, 0}_{n - \log s + 1})$-tree $T$ and a 
\newline
$(0, \underbrace{2, \ldots, 2}_{\log s - r - 4}, 0, \underbrace{4, \ldots, 4}_{\lceil \frac{r}{2} \rceil + 2 \lfloor \frac{r}{2} \rfloor - \frac{n}{2}}, 0, \ldots, 0)$-tree $T'$. (Note that the term $\frac{n}{2} - 2 \lfloor \frac{r}{2} \rfloor$ is nonnegative.) Note that Proposition \ref{prop:NeighborConjeHelpProp4} guarantees the existence of $T$. So it remains to show that we can obtain $T'$. By Proposition \ref{prop:NeighborConjeHelpProp2} this can be reduced to showing the existence of a 
\newline
$(\underbrace{1, \ldots, 1}_{\frac{n}{2} - \lfloor \frac{r}{2} \rfloor}, 0, \underbrace{2, \ldots, 2}_{\log s - r - 4}, 0, \underbrace{4, \ldots, 4}_{\lceil \frac{r}{2} \rceil + 2 \lfloor \frac{r}{2} \rfloor - \frac{n}{2}}, \underbrace{0, \ldots, 0}_{n - \log s + 2})$-tree $T''$. Due to Proposition  \ref{prop:NeighborConjeHelpProp4} $T''$ can be obtained. \hfill $\qed$

\emph{Proof of Proposition \ref{prop:mainprop2}:} Due to Proposition \ref{prop:NeighborConjeHelpProp2} it suffices to show that there is a 
\newline
$(\underbrace{1, \ldots, 1}_{r - j - 1}, 0, \underbrace{2, \ldots, 2}_{r - j}, 0, \underbrace{4, \ldots, 4}_{j}, 0, \ldots, 0)$-tree. (Note that $(r - j + 1) + (r - j) + j = 2r - j + 1 \leq \log s - 1$.) By assumption there is a
\newline
$(0, \underbrace{2, \ldots, 2}_{r - j - 1}, 0, \underbrace{4, \ldots, 4}_{j + 1}, 0, \ldots, 0)$-tree. So by Proposition \ref{prop:NeighborConjeHelpProp3} it suffices to show that we can obtain a 
\newline $(0, \underbrace{0, \ldots, 0}_{r - j - 1}, 0, \underbrace{0, \ldots, 0}_{j + 1}, \underbrace{4, \ldots, 4}_{r - 2j - 1}, 0, \underbrace{8, \ldots, 8}_{j}, 0, \ldots, 0)$-tree. (Note that $r - 2j - 1 > 0$.) 
We distinguish two cases.
\begin{list}{}{\leftmargin=20pt} 
\item[Case 1] $j \leq \frac{n}{8}$. According to Proposition \ref{prop:NeighborConjeHelpProp3} we are left with proving the existence of both a 
\newline
$(\underbrace{0, \ldots, 0}_{r + 3}, \underbrace{0, \ldots, 0}_{r - \frac{n}{4} - 1}, \underbrace{8, \ldots, 8}_{\frac{n}{4} - 2j}, 0, \underbrace{16, \ldots, 16}_{j}, 0, \ldots, 0)$-tree $T$ and a 
$(\underbrace{0, \ldots, 0}_{r + 3}, \underbrace{8, \ldots, 8}_{r - \frac{n}{4} - 1}, 0, \ldots, 0)$-tree $T'$.
\newline
We observe that in the sequence corresponding to $T$ all but the first $\log s$ entries are zero. 
(This can be seen by distinguishing the cases $j > 0$ and $j = 0$.)
So we can apply
Proposition \ref{prop:NeighborConjeHelpProp4}, which guarantees the existence of $T$. To show that we can obtain $T'$ it suffices by Proposition \ref{prop:NeighborConjeHelpProp2} to prove that there is a 
$(\underbrace{1, \ldots, 1}_{4n - 8r + 8}, \underbrace{0, \ldots, 0}_{r + 3}, \underbrace{8, \ldots, 8}_{r - \frac{n}{4} - 1}, 0, \ldots, 0)$-tree. (Note that $(4n - 8r + 8) +  (r + 3) + (r - \frac{n}{4} - 1) \leq
\frac{15}{4}n - 6r + 10 \leq \frac{3}{4} n + O(\log n) \leq \log s$.) Due to Proposition \ref{prop:NeighborConjeHelpProp4} such a tree exists.
\item[Case 2] $j \geq \frac{n}{8}$.
According to Proposition \ref{prop:NeighborConjeHelpProp3} we are left with proving the existence of both a 
\newline
$(\underbrace{0, \ldots, 0}_{r + 3}, \underbrace{0, \ldots, 0}_{r - 2j - 1}, 0, \underbrace{0, \ldots, 0}_{j- \frac{n}{8}}, \underbrace{16, \ldots, 16}_{\frac{n}{8}}, 0, \ldots, 0)$-tree $T$ and a 
\newline
$(\underbrace{0, \ldots, 0}_{r + 3}, \underbrace{8, \ldots, 8}_{r - 2j - 1}, 0, \underbrace{16, \ldots, 16}_{j - \frac{n}{8}}, 0, \ldots, 0)$-tree $T'$
\newline
Proposition \ref{prop:NeighborConjeHelpProp4} guarantees the existence of $T$. To show that we can obtain $T'$ it suffices by Proposition \ref{prop:NeighborConjeHelpProp2} to prove that there is a 
\newline
$(\underbrace{1, \ldots, 1}_{\frac{n}{8}}, \underbrace{0, \ldots, 0}_{r + 3}, \underbrace{8, \ldots, 8}_{r - 2j - 1},0, \underbrace{16, \ldots, 16}_{j - \frac{n}{8}}, 0, \ldots, 0)$-tree. Due to Proposition \ref{prop:NeighborConjeHelpProp4} such a tree exists. \hfill $\qed$
\end{list}

\emph{Proof of Proposition \ref{prop:mainprop3}:} Note that by assumption $i \leq r - 1 \leq 
\frac{\log s - 4}{2}$.
By Proposition \ref{prop:NeighborConjeHelpProp2} it suffices to show that there is a 
\newline
$(\underbrace{1, \ldots, 1}_{\log s - i - 3}, 0, \underbrace{2, \ldots, 2}_{i}, 0, \ldots, 0)$-tree. By the fact that there is a
\newline
$(0, \underbrace{2, \ldots, 2}_{i + 1}, 0, \ldots, 0)$-tree and Proposition \ref{prop:NeighborConjeHelpProp3} we are left with showing that there is a 
\newline
$(\underbrace{0, \ldots, 0}_{i + 2}, \underbrace{2, \ldots, 2}_{\log s - 2i - 4}, 0, \underbrace{4, \ldots, 4}_{i}, 0, \ldots, 0)$-tree.
We distinguish two cases.
\begin{list}{}{\leftmargin=20pt} 
\item[Case 1] $i \leq \frac{n}{4}$. By Proposition \ref{prop:NeighborConjeHelpProp3} it suffices to show that there is both a 
\newline
$(\underbrace{0, \ldots, 0}_{i + 3}, \underbrace{0, \ldots, 0}_{\log s - \frac{n}{2} - 4}, \underbrace{4, \ldots, 4}_{\frac{n}{2} - 2i}, 0, \underbrace{8, \ldots, 8}_{i}, 0, \ldots, 0)$-tree $T$ and a 
\newline
$(\underbrace{0, \ldots, 0}_{i + 3}, \underbrace{4, \ldots, 4}_{\log s - \frac{n}{2} - 4}, 0, \ldots, 0)$-tree $T'$. Proposition \ref{prop:NeighborConjeHelpProp4} guarantees that $T$ exists. It remains to show that $T'$ can be obtained. According to Proposition \ref{prop:NeighborConjeHelpProp2} it suffices to show that there is a 
\newline
$(\underbrace{1, \ldots, 1}_{\frac{n}{2} - i + 1}, \underbrace{0, \ldots, 0}_{i + 3}, \underbrace{4, \ldots, 4}_{\log s - \frac{n}{2} - 4}, 0, \ldots, 0)$-tree, which exists due to Proposition  \ref{prop:NeighborConjeHelpProp4}.
\item[Case 2] $i \geq \frac{n}{4}$. Due to Proposition \ref{prop:NeighborConjeHelpProp3} we are left with showing that there is both a 
\newline
$(\underbrace{0, \ldots, 0}_{i + 3}, \underbrace{0, \ldots, 0}_{\log s - 2i - 4}, 0, \underbrace{0, \ldots, 0}_{i - \frac{n}{4}}, \underbrace{8, \ldots, 8}_{\frac{n}{4}}, 0, \ldots, 0)$-tree $T$ and a 
\newline
$(\underbrace{0, \ldots, 0}_{i + 3}, \underbrace{4, \ldots, 4}_{\log s - 2i - 4}, 0, \underbrace{8, \ldots, 8}_{i - \frac{n}{4}}, 0, \ldots, 0)$-tree $T'$. Due to Proposition \ref{prop:NeighborConjeHelpProp4} $T$ exists. To show that $T'$ can be obtained it suffices to prove that there is a 
\newline
$(\underbrace{1, \ldots, 1}_{\frac{n}{4}}, \underbrace{0, \ldots, 0}_{i + 3}, 
\underbrace{4, \ldots, 4}_{\log s - 2i - 4}, 0, \underbrace{8, \ldots 8}_{i - \frac{n}{4}}, 0, \ldots, 0)$-tree, which exists due to Proposition \ref{prop:NeighborConjeHelpProp4}. \hfill $\qed$
\end{list}

\subsection{Proof of Theorem  \ref{theo:NeighbourhoodPartbnotworkforsomeotherc}}

We consider the proof of Theorem \ref{theo:NeighbourhoodPartbnotworkc1},
 plug in $s := \frac{2^{n - 1}}{cn}$ (instead of $s := \frac{2^{n - 1}}{n}$) and modify the proofs of Proposition \ref{prop:mainprop1} - \ref{prop:mainprop3}. This will form our proof of Theorem  \ref{theo:NeighbourhoodPartbnotworkforsomeotherc}. 
So it remains to adapt the proofs of Proposition \ref{prop:mainprop1} - \ref{prop:mainprop3}.

\emph{Proof of Proposition \ref{prop:mainprop1}:} 
 By Proposition \ref{prop:NeighborConjeHelpProp2} it suffices to show that there is a 
\newline
$(\underbrace{1, \ldots, 1}_{\log s - r - 4}, 0, \underbrace{2, \ldots, 2}_{\lceil \frac{r}{2} \rceil}, 0, \underbrace{4, \ldots, 4}_{\lfloor \frac{r}{2} \rfloor}, \underbrace{0, \ldots, 0}_{n - \log s + 2})$-tree. According to Proposition \ref{prop:NeighborConjeHelpProp3} it suffices to show that there exists both a 
\newline
$(\underbrace{0, \ldots, 0}_{\log s - r - 3}, 0, \underbrace{0, \ldots, 0}_{\lceil \frac{r}{2} \rceil + 2 \lfloor \frac{r}{2} \rfloor - \frac{cn}{2}}, \underbrace{4, \ldots, 4}_{\frac{cn}{2} - 2 \lfloor \frac{r}{2} \rfloor} , 0, \underbrace{8, \ldots, 8}_{\lfloor \frac{r}{2} \rfloor}, \underbrace{0, \ldots, 0}_{n - \log s + 1})$-tree $T$ and a 
\newline
$(0, \underbrace{2, \ldots, 2}_{\log s - r - 4}, 0, \underbrace{4, \ldots, 4}_{\lceil \frac{r}{2} \rceil + 2 \lfloor \frac{r}{2} \rfloor - \frac{cn}{2}}, 0, \ldots, 0)$-tree $T'$. (Note that the term $\frac{cn}{2} - 2 \lfloor \frac{r}{2} \rfloor$ is nonnegative.) Note that Proposition \ref{prop:NeighborConjeHelpProp4} guarantees the existence of $T$. So it remains to show that we can obtain $T'$. By Proposition \ref{prop:NeighborConjeHelpProp2} this can be reduced to showing the existence of a 
\newline
$(\underbrace{1, \ldots, 1}_{\frac{cn}{2} - \lfloor \frac{r}{2} \rfloor}, 0, \underbrace{2, \ldots, 2}_{\log s - r - 4}, 0, \underbrace{4, \ldots, 4}_{\lceil \frac{r}{2} \rceil + 2 \lfloor \frac{r}{2} \rfloor - \frac{cn}{2}}, \underbrace{0, \ldots, 0}_{n - \log s + 2})$-tree $T''$. 
Let $t := \frac{cn}{2} - \lfloor \frac{r}{2} \rfloor + 2 \cdot (\log s - r - 4) + 4 \cdot (\lceil \frac{r}{2} \rceil + 2 \lfloor \frac{r}{2} \rfloor - \frac{cn}{2})$.
We have $t \geq \frac{15}{4}n - \frac{3}{2}cn + O(\log n) \geq 2cn$ (since $c < \frac{15}{14}$).
So by Proposition  \ref{prop:NeighborConjeHelpProp4} $T''$ can be obtained. 
\hfill $\qed$

\emph{Proof of Proposition \ref{prop:mainprop2}:}
Due to Proposition \ref{prop:NeighborConjeHelpProp2} it suffices to show that there is a 
\newline
$(\underbrace{1, \ldots, 1}_{r - j - 1}, 0, \underbrace{2, \ldots, 2}_{r - j}, 0, \underbrace{4, \ldots, 4}_{j}, 0, \ldots, 0)$-tree. (Note that $(r - j + 1) + (r - j) + j = 2r - j + 1 \leq \log s - 1$.) By assumption there is a
\newline
$(0, \underbrace{2, \ldots, 2}_{r - j - 1}, 0, \underbrace{4, \ldots, 4}_{j + 1}, 0, \ldots, 0)$-tree. So by Proposition \ref{prop:NeighborConjeHelpProp3} it suffices to show that we can obtain a 
\newline $(0, \underbrace{0, \ldots, 0}_{r - j - 1}, 0, \underbrace{0, \ldots, 0}_{j + 1}, \underbrace{4, \ldots, 4}_{r - 2j - 1}, 0, \underbrace{8, \ldots, 8}_{j}, 0, \ldots, 0)$-tree. (Note that $r - 2j - 1 > 0$.) 
We distinguish two cases.
\begin{list}{}{\leftmargin=20pt} 
\item[Case 1] $j \leq \frac{cn}{8}$. According to Proposition \ref{prop:NeighborConjeHelpProp3} we are left with proving the existence of both a 
\newline
$(\underbrace{0, \ldots, 0}_{r + 3}, \underbrace{0, \ldots, 0}_{r - \frac{cn}{4} - 1}, \underbrace{8, \ldots, 8}_{\frac{cn}{4} - 2j}, 0, \underbrace{16, \ldots, 16}_{j}, 0, \ldots, 0)$-tree $T$ and a 
$(\underbrace{0, \ldots, 0}_{r + 3}, \underbrace{8, \ldots, 8}_{r - \frac{cn}{4} - 1}, 0, \ldots, 0)$-tree $T'$.
\newline
We observe that in the sequence corresponding to $T$ all but the first $\log s$ entries are zero. 
(This can be seen by distinguishing the cases $j > 0$ and $j = 0$.)
So we can apply
Proposition \ref{prop:NeighborConjeHelpProp4}, which guarantees the existence of $T$. To show that we can obtain $T'$ it suffices by Proposition \ref{prop:NeighborConjeHelpProp2} to prove that there is a 
$(\underbrace{1, \ldots, 1}_{4cn - 8r + 8}, \underbrace{0, \ldots, 0}_{r + 3}, \underbrace{8, \ldots, 8}_{r - \frac{cn}{4} - 1}, 0, \ldots, 0)$-tree. (Note that $(4cn - 8r + 8) +  (r + 3) + (r - \frac{cn}{4} - 1) \leq
\frac{15}{4}cn - 6r + 10 \leq (1 - \epsilon)n + O(\log n)$ for some constant $\epsilon > 0$; additionally 
$(1 - \epsilon)n + O(\log n) \leq \log s$.) Due to Proposition \ref{prop:NeighborConjeHelpProp4} such a tree exists.
\item[Case 2] $j \geq \frac{cn}{8}$.
According to Proposition \ref{prop:NeighborConjeHelpProp3} we are left with proving the existence of both a 
\newline
$(\underbrace{0, \ldots, 0}_{r + 3}, \underbrace{0, \ldots, 0}_{r - 2j - 1}, 0, \underbrace{0, \ldots, 0}_{j- \frac{cn}{8}}, \underbrace{16, \ldots, 16}_{\frac{cn}{8}}, 0, \ldots, 0)$-tree $T$ and a 
\newline
$(\underbrace{0, \ldots, 0}_{r + 3}, \underbrace{8, \ldots, 8}_{r - 2j - 1}, 0, \underbrace{16, \ldots, 16}_{j - \frac{cn}{8}}, 0, \ldots, 0)$-tree $T'$
\newline
Proposition \ref{prop:NeighborConjeHelpProp4} guarantees the existence of $T$. To show that we can obtain $T'$ it suffices by Proposition \ref{prop:NeighborConjeHelpProp2} to prove that there is a 
\newline
$(\underbrace{1, \ldots, 1}_{\frac{cn}{8}}, \underbrace{0, \ldots, 0}_{r + 3}, \underbrace{8, \ldots, 8}_{r - 2j - 1},0, \underbrace{16, \ldots, 16}_{j - \frac{cn}{8}}, 0, \ldots, 0)$-tree. Due to Proposition \ref{prop:NeighborConjeHelpProp4} such a tree exists. Indeed, $\frac{cn}{8} + 8(r - 2j - 1) + 16(j - \frac{cn}{8}) = 4n - \frac{15}{8}cn + O(\log n) \geq 2cn$ (since $c < \frac{32}{31}$).
\hfill $\qed$
\end{list}

\emph{Proof of Proposition \ref{prop:mainprop3}:}
Note that by assumption $i \leq r - 1 \leq 
\frac{\log s - 4}{2}$.
By Proposition \ref{prop:NeighborConjeHelpProp2} it suffices to show that there is a 
\newline
$(\underbrace{1, \ldots, 1}_{\log s - i - 3}, 0, \underbrace{2, \ldots, 2}_{i}, 0, \ldots, 0)$-tree. By the fact that there is a
\newline
$(0, \underbrace{2, \ldots, 2}_{i + 1}, 0, \ldots, 0)$-tree and Proposition \ref{prop:NeighborConjeHelpProp3} we are left with showing that there is a 
\newline
$(\underbrace{0, \ldots, 0}_{i + 2}, \underbrace{2, \ldots, 2}_{\log s - 2i - 4}, 0, \underbrace{4, \ldots, 4}_{i}, 0, \ldots, 0)$-tree.
We distinguish two cases.
\begin{list}{}{\leftmargin=20pt} 
\item[Case 1] $i \leq \frac{cn}{4}$. By Proposition \ref{prop:NeighborConjeHelpProp3} it suffices to show that there is both a 
\newline
$(\underbrace{0, \ldots, 0}_{i + 3}, \underbrace{0, \ldots, 0}_{\log s - \frac{cn}{2} - 4}, \underbrace{4, \ldots, 4}_{\frac{cn}{2} - 2i}, 0, \underbrace{8, \ldots, 8}_{i}, 0, \ldots, 0)$-tree $T$ and a 
\newline
$(\underbrace{0, \ldots, 0}_{i + 3}, \underbrace{4, \ldots, 4}_{\log s - \frac{cn}{2} - 4}, 0, \ldots, 0)$-tree $T'$. Proposition \ref{prop:NeighborConjeHelpProp4} guarantees that $T$ exists. It remains to show that $T'$ can be obtained. According to Proposition \ref{prop:NeighborConjeHelpProp2} it suffices to show that there is a 
\newline
$(\underbrace{1, \ldots, 1}_{\frac{cn}{2} - i}, \underbrace{0, \ldots, 0}_{i + 3}, \underbrace{4, \ldots, 4}_{\log s - \frac{cn}{2} - 4}, 0, \ldots, 0)$-tree $T''$.
We have $\frac{cn}{2} - i + 4 \cdot (\log s - \frac{cn}{2} - 4) \geq 4n - \frac{7}{4}cn + O(\log n) \geq 2cn$ (since $c < \frac{15}{16}$) and so Proposition  \ref{prop:NeighborConjeHelpProp4} proves that $T''$ exists.
\item[Case 2] $i \geq \frac{cn}{4}$. Due to Proposition \ref{prop:NeighborConjeHelpProp3} we are left with showing that there is both a 
\newline
$(\underbrace{0, \ldots, 0}_{i + 3}, \underbrace{0, \ldots, 0}_{\log s - 2i - 4}, 0, \underbrace{0, \ldots, 0}_{i - \frac{cn}{4}}, \underbrace{8, \ldots, 8}_{\frac{cn}{4}}, 0, \ldots, 0)$-tree $T$ and a 
\newline
$(\underbrace{0, \ldots, 0}_{i + 3}, \underbrace{4, \ldots, 4}_{\log s - 2i - 4}, 0, \underbrace{8, \ldots, 8}_{i - \frac{cn}{4}}, 0, \ldots, 0)$-tree $T'$. Due to Proposition \ref{prop:NeighborConjeHelpProp4} $T$ exists. To show that $T'$ can be obtained it suffices to prove that there is a 
\newline
$(\underbrace{1, \ldots, 1}_{\frac{cn}{4}}, \underbrace{0, \ldots, 0}_{i + 3}, 
\underbrace{4, \ldots, 4}_{\log s - 2i - 4}, 0, \underbrace{8, \ldots 8}_{i - \frac{cn}{4}}, 0, \ldots, 0)$-tree $T''$.
\newline
We have $\frac{cn}{4} + 4 \cdot(\log s - 2i - 4) + 8 \cdot (i - \frac{cn}{4}) \geq 4n - \frac{7}{4}cn + O(\log n) \geq 2cn$ (since $c < \frac{16}{15}$) and therefore Proposition \ref{prop:NeighborConjeHelpProp4} guarantees the existence of $T''$.
\hfill $\qed$
\end{list}

\section{Constructing unsatisfiable $k$-CNF formulas with small neighborhood} \label{sec:construunsatformusmanei}

\emph{Proof of Theorem \ref{theo:UnsatCNFOnlyPowers}:} Theorem \ref{theo:NeighbourhoodPartbnotworkc1} and Theorem \ref{theo:equivalence} directly imply Theorem \ref{theo:UnsatCNFOnlyPowers}. $\Box$

Let $\cal{F}$ be a hypergraph. We say that Maker uses a \emph{pure pairing strategy} if at the beginning of the game he divides all but at most one of the vertices of $\cal{F}$ into pairs, lets Breaker start the game, and whenever Breaker claims one vertex of a pair he takes the other one. 
\begin{obse} \label{obse:pairingandpurepairing}
If there is a $(k,s)$-hypergraph $\cG$ then there is a $k$-uniform hypergraph with maximum degree at most $s$ where Maker has a winning pure pairing strategy.
\end{obse}
\noindent
This can be seen as follows. Let $\cG$ be a $(k,s)$-hypergraph, let ${\cG}'$ be a disjoint copy of $\cG$ and let $\cH$ be the hypergraph with $V({\cH}) = V({\cG}) \cup V({\cG}')$ and $E({\cH}) = E({\cG}) \cup E({\cG}')$. 
Clearly, $\cH$ is a $k$-uniform hypergraph with maximum degree at most $s$.
Moreover,  let $S$ and $S'$ denote the winning pairing strategy of Maker in $\cG$ and ${\cG}'$, respectively, and let $v_{S}$ and $v_{S'}$ denote the corresponding vertices Maker claims in the first round. We consider the pure pairing strategy $\tilde{S}$ where the pairings corresponding to $S$ and $S'$ are maintained and additionally $v_{S}$ is paired with $v_{S'}$.
Clearly $\tilde{S}$ is a winning pure pairing strategy for Maker. Indeed, it allows him to play his original strategy in at least one of the hypergraphs ${\cG}, {\cG}'$, which implies that at the end Maker owns a full hyperedge of \nolinebreak $\cH$.

\emph{Proof of Theorem \ref{theo:equivalence}:} We first show (i). 
Due to Observation \ref{obse:pairingandpurepairing} we can assume that there is a $k$-uniform hypergraph $\cG$ with maximum degree at most $s$ where Maker has a winning pure pairing strategy $S$. Let $(v_{1}, v'_{1}), (v_{2}, v'_{2}), \ldots, (v_{r}, v'_{r})$ be the pairing of $V(\cG)$ corresponding to $S$. To construct an unsatisfiable balanced $(k,2s)$-CNF formula we proceed as follows. First we form for every hyperedge $e = (w_{1}, w_{2}, \ldots, w_{k})$ of $\cG$ a clause $\cC_{e} = (w_{1} \vee w_{2} \vee \ldots \vee w_{k})$ and set ${\cal{F}} := \wedge_{e \in E(\cG)} {\cC}_{e}$ with $ E(\cG)$ denoting the hyperedge set of $\cal{G}$.
Then we replace (in $\cal{F}$) $v_{i}$ and $v'_{i}$ with $x_{i}$ and $\bar{x}_{i}$, respectively, for every $i$, $i = 1, \ldots, r$. Note that by construction every literal $x \in \{x_{i}, \bar{x}_{i}\}$ occurs in at most $s$ clauses of $\cal{F}$.
It remains to show that $\cal{F}$ is unsatisfiable. 

Note that by playing according to $S$ Maker achieves that the outcome of the game corresponds to a valid assignment of $\cal{F}$ with
\begin{displaymath}
x_{i} = \left\{ 
\begin{array}{ll}
\text{true}, & \text{if Breaker claims $v_{i}$} \\
\text{false}, &  \text{if Breaker claims $v'_{i}$}
\end{array} \right.
\end{displaymath}
Due to our construction $\cal{F}$ is satisfiable if and only if Breaker has a winning strategy in $\cG$ against $S$. But by assumption $S$ is a winning strategy for Maker, implying that $\cal{F}$ is not satisfiable.
It remains to prove (ii). 
Let $\cal{F}$ be an unsatisfiable $(k,s)$-CNF formula and let $\{x_{1}, \ldots, x_{r}\}$ be the set of variables of $\cal{F}$. To construct a $(k,s)$-hypergraph we proceed as follows. For every clause $\cC = (x_{1} \vee x_{2} \vee \ldots \vee x_{k})$ of $\cal{F}$ we construct a hyperedge $e_{\cC} = (x_{1}, x_{2}, \ldots, x_{k})$ and let $\cG$ be the hypergraph with vertex set $\{x_{1}, \ldots, x_{r}\} \cup \{\bar{x}_{1}, \ldots, \bar{x}_{r}  \}$ and hyperedge set $\{e_{\cC} : \text{ $\cC$ is a clause of \cal{F}}\}$. 
We denote by $S$ the pure pairing strategy where $x_{i}$ is paired with $\bar{x}_{i}$ for every $i$, $i = 1, \ldots r$. Similarly as above we get that Breaker has a winning strategy against $S$ if and only if $\cal{F}$ is satisfiable. Hence Maker has a winning (pure) pairing strategy on $\cG$ and therefore $\cal{G}$ is a $(k,s)$-hypergraph. \hfill $\qed$

\emph{Proof of Corollary \ref{coro:UnsatCNFimpliesf1}:}
Davydov, Davydova, and Kleine B\"{u}ning  \cite{DDK} established the following characterization for MU(1)-formulas. ($\vbl(F)$ denotes the set of variables which occur in the formula $F$.)
\begin{lemm} \label{lemm:DDK} {\bf(Davydov, Davydova, and Kleine B\"{u}ning  \cite{DDK})} $F \in \text{MU(1)}$ if and only if either $F = \{ \emptyset \}$ or $F$ is the disjoint union of formulas $F'_{1}, F'_{2}$ such that for a variable $x$ we have
\begin{itemize}
\item $\vbl(F'_{1}) \cap \vbl(F'_{2}) = \{x\}$ and $\{x, \bar{x}\} \subseteq \bigcup_{C \in F} C$;
\item $F_{1} := \{C \backslash \{x\}: C \in F'_{1}\} \in \text{MU(1)}$;
\item $F_{2} := \{C \backslash \{\bar{x}\}: C \in F'_{2}\} \in \text{MU(1)}$.
\end{itemize}
\end{lemm} 
\noindent
The proofs of 
Theorem \ref{theo:NeighbourhoodPartbnotworkc1}, Theorem \ref{theo:UnsatCNFOnlyPowers}, and  Theorem \ref{theo:equivalence} implicitly yield an unsatisfiable $(k, \frac{2^{k}}{k})$-CNF formula $\cal{F}$ (for sufficiently large $k$ which are a power of 2) and an unsatisfiable $(k, 2 \cdot \frac{2^{k}}{k})$-CNF formula ${\cal{F}}'$ (for sufficiently large $k$). It can be seen that $\cal{F}$ and ${\cal{F}}'$ have the properties stated in Lemma \ref{lemm:DDK}, implying that they both belong to MU(1). $\Box$
%

\emph{Proof of Theorem \ref{theo:neighborhoodbounds}:} Part (ii) follows directly from the construction used in the proof of Lemma \ref{lemm:NeighbourhoodPartaAuxilLemm}. (By Theorem \ref{theo:equivalence} we can interpret the corresponding tree as a boolean formula $\cal{F}$. Carefully counting the maximum neighborhood size then shows that $\cal{F}$ is a $(k, 2^{k - 1} + 2^{k - 2} + 1)$-CNF). It remains to prove part (i). This is an immediate consequence of 
Lemma \ref{lemm:constructioninvolvedtreereduction}. Indeed, let $\cal{F}$ be the boolean formula corresponding to the tree guaranteed by Lemma \ref{lemm:constructioninvolvedtreereduction}. Note that $\cal{F}$ has the property that two neighboring clauses $C,D$ of $\cal{F}$ always form a conflict. This implies that the neighborhood size of a clause  $(x_{1} \vee x_{2} \vee \ldots \vee x_{k})$ is bounded by $\sum_{i = 1}^{k} d(\bar{x_{i}})$ with $d(\bar{x_{i}})$ denoting the number of occurrences of $\bar{x_{i}}$.
Moreover, by construction the boolean formula $\cal{F}$ corresponding to the tree guaranteed by Lemma \ref{lemm:constructioninvolvedtreereduction} has the property that every literal occurs in at most $\frac{2^{k - 1}}{k}$ clauses, implying that the maximum neighbourhood size of $\cal{F}$ is at most $k \cdot \frac{2^{k - 1}}{k} = 2^{k - 1}$. \hfill $\qed$

\emph{Proof of Theorem \ref{theo:improvedbounds}:} Along similar lines as above we can show that Theorem \ref{theo:NeighbourhoodPartbnotworkforsomeotherc} implies that $f(k) \leq \frac{2^{k - 1}}{ck} - 1$ and
$l(k) \leq \frac{2^{k-1}}{c} - 1$. \hfill $\qed$

\textbf{Acknowledgment:} We would like to thank Tibor Szab\'o for the intensive support and Emo Welzl for the numerous helpful remarks.



\section*{Appendix}

 \begin{appendix}

\section{Establishing a proper halving 2-coloring}

\emph{Proof of Theorem \ref{theo:NeighbourhoodPartProperHalvingIsOk}:} For simplicity we only consider hypergraphs with an even number of vertices. We will show the following stronger claim. 
\begin{prop} \label{prop:NeighbourhoodPartasworksforhalvingstronger}
Let $\cal{F}$ be a $n$-uniform hypergraph with $2r$ vertices and maximum degree at most $\frac{2^{n - 2}}{en}$. Then for each pairing $(v_{1}, v'_{1}), (v_{2}, v'_{2}), \ldots, (v_{r}, v'_{r})$ of $V(\cal{F})$ there is a proper 2-coloring such that $v_{i}$ and $v'_{i}$ have different colors for every $i$, $i = 1, \ldots, r$.
\end{prop}
Before starting with the proof we need some notation. Let $P = (v_{1}, v'_{1}), (v_{2}, v'_{2}), \ldots, (v_{r}, v'_{r})$ be a pairing of $V(\cal{F})$. By a \emph{(proper) $P$-2-coloring} we denote a (proper) 2-coloring of $V(\cal{F})$ such that $v_{i}$ and $v'_{i}$ have different colors for every $k$, $k = 1, \ldots, r$.  
Moreover, for every vertex $x \in V(\cal{F})$ we denote by $f(x)$ the vertex which is paired with $x$ in $P$ (i.e., $f(v_{i}) = v'_{i}$ and $f(v'{i}) = v_{i}$).
\newline
\emph{Proof of Proposition \ref{prop:NeighbourhoodPartasworksforhalvingstronger}:} 
Suppose, for a contradiction, that there is a pairing 
\newline
$P = (v_{1}, v'_{1}), (v_{2}, v'_{2}), \ldots, (v_{r}, v'_{r})$ of $V(\cal{F})$ such that there is no proper $P$-2-coloring. For every hyperedge $e = (x_{1}, x_{2} \ldots, x_{n})$ we add the hyperedge $e' = (f(x_{1}), f(x_{2}) \ldots, f(x_{r}))$ to $\cal{F}$ and denote the resulting hypergraph by ${\cal{F}}'$. Note that $\Delta({\cal{F}}') \leq  2 \cdot \Delta({\cal{F}}) \leq \frac{2^{n - 1}}{en}$. By construction, every $P$-2-coloring of $V({\cal{F}}')$ has both a monochromatic red hyperedge and a  monochromatic blue hyperedge. Hence, if Maker plays the pairing strategy corresponding to $P$ he can completely occupy some hyperedge by the end of the game. So there is an $(n, \frac{2^{n - 1}}{en})$-hypergraph. Due to Theorem \ref{theo:equivalence} there is an unsatisfiable $(n, \frac{2^{n}}{en})$-CNF formula, which contradicts Theorem \ref{theo:KSTlowerbound}. $\Box$

\end{appendix}

\end{document}